\documentclass[twocolumn,showpacs,amsmath,amssymb,prb]{revtex4-2}

\usepackage{natbib}

\usepackage[version=4]{mhchem}
\usepackage{physics}
\usepackage{siunitx}
\usepackage{graphicx}
\usepackage[table]{xcolor}
\usepackage{lipsum}
\usepackage[T1]{fontenc}
\usepackage[format=hang,justification=centerlast,font={small,stretch=1},labelfont=bf,labelsep=colon]{caption}
\usepackage{subcaption}
\usepackage[backref=true,colorlinks=true,citecolor=blue,urlcolor=blue,linkcolor=blue]{hyperref}

\begin{document}
\sloppy

\title{\textbf{Non-linear Hall resistivity in overdoped \ce{Pr_{1.82}Ce_{0.18}CuO_{4 + \delta}} electron-doped cuprate}}

\author{S. Ghotb}
\author{M. Dion}
\author{G. Hardy}
\author{P. Fournier}
\email[Corresponding author: ]{patrick.fournier@usherbrooke.ca}
\affiliation{Institut Quantique, Regroupement qu\'{e}b\'{e}cois sur les mat\'{e}riaux de pointe, D\'{e}partement de Physique, Universit\'{e} de Sherbrooke, Sherbrooke, Qu\'{e}bec, Canada, J1K 2R1}

\begin{abstract}
The Hall coefficient $R_\mathrm{H}$ as a function of temperature for \ce{Pr_{2-x}Ce_{x}CuO_{4}} (\ce{PCCO}) at $x = 0.17$ shows two sign changes between \SIlist[list-units=single]{0;300}{\kelvin} that can be explained qualitatively using the two-carrier model. Using this two-carrier model,  one predicts the presence of a nonlinear $B^{3}$ contribution to the Hall resistivity $\rho_\mathrm{yx}$ that should be easily observed when the linear coefficient $R_\mathrm{H}$ is approaching zero. We present the measurement of this nonlinear Hall resistivity for overdoped \ce{PCCO} thin films at $x=0.18$. We show that this weak nonlinear term persists at all temperatures despite being dominated by the linear term ($R_\mathrm{H}B$). It is strongly temperature dependent and, analyzing it with the two-carrier model, we conclude that the density of electron carriers is larger than the hole density, but that the hole mobility is larger than that of the electrons. This nonlinear Hall resistivity should be observable for similar materials for which $R_\mathrm{H}(T)$ goes through sign reversals as does overdoped \ce{PCCO}.
\end{abstract}

\maketitle

\section{Introduction}

Driven by strong electron-electron interactions, the undoped parent compounds of cuprate superconductors are antiferromagnetic Mott insulators. Charge carrier doping with holes or electrons into their copper-oxygen (\ce{CuO_{2}}) planes is usually achieved by chemical substitution. It leads to the suppression of their insulating phase and antiferromagnetism,  and to the emergence of a strange metallic phase exhibiting superconductivity \cite{armitage2010progress,orenstein2000advances}. A debate still persists over the nature of this peculiar metallic phase where anomalies such as the pseudogap are affecting strongly the normal state properties \cite{armitage2010progress,matsui2005angle,doiron2017pseudogap,greene2020strange,hussey2008phenomenology}. Unlocking the mysteries behind this normal state strange metal is a priority in our goal to understand the origin of superconductivity in cuprates and related materials.

The Hall coefficient, $R_\mathrm{H}$, in electron-doped cuprates of general formula \ce{R_{2-x}Ce_{x}CuO_{4}} (R = Pr, Nd, Sm, La,. . .) shows a nontrivial behavior as a function of doping and temperature as strong electron correlations reshape their electronic structure over the whole temperature-doping ($T-x$) phase diagram \cite{armitage2010progress,fournier2015t}. At zero temperature, the doping dependence of $R_\mathrm{H}$ results from a multistep transformation (reconstruction) of the Fermi surface (FS) \cite{dagan2016fermi,charpentier2010antiferromagnetic} from unconnected Fermi arcs at low doping to a large hole cylinder at large doping consistent with the observations from angle-resolved photoemission spectroscopy (ARPES)  \cite{matsui2007evolution,armitage2002doping,helm2009evolution,he2019fermi,helm2010magnetic,kartsovnik2011fermi}. At low doping ($x \lesssim 0.12$), a negative $R_\mathrm{H}(0)$ decreases in magnitude roughly as $R_\mathrm{H}(0) \sim -1 / x$ as only electronic states on Fermi arcs in proximity to the $\vec{k}=(\pm \pi,0)$ and $\vec{k}=(0,\pm \pi)$ antinodal points in the Brillouin zone (BZ) contribute to electrical transport. At large doping, beyond $x^{*} \cong 0.165$, a positive $R_\mathrm{H}(0)$ increases as $R_\mathrm{H}(0) \sim 1 / (1-x)$ tracking closely the expected contribution from a slowly-filling large hole like cylindrical FS centered around $\vec{k} = (\pi,\pi)$ as also observed by ARPES \cite{matsui2007evolution}. 

Through the transition doping range $0.12 \leq x \leq 0.17$, $R_\mathrm{H}(0)$ changes sign rapidly with $x$, bridging the gap between the low and high doping regimes \cite{dagan2016fermi,charpentier2010antiferromagnetic}. In this specific doping range, ARPES confirms the presence of two sets of disconnected Fermi arcs separated by the pseudogap at the so-called hot spots. These arcs include the antinodal ones already observed at low doping and another set close to $\vec{k}=(\pm \pi,\pm \pi)$ in the nodal directions. Moreover, apparent band folding also observed by ARPES \cite{he2019fermi} may indicate the presence of closed electron and hole pockets in this doping range. As a consequence, a two-carrier model is often used to explain qualitatively the doping and temperature dependence of $R_\mathrm{H}$ in this transition doping range ascribing the antinodal Fermi arcs to electron-like carriers and the nodal ones to hole like carriers \cite{sebastian2015quantum,hozoi2008fermiology,galitski2009paired,he2019fermi}.   

\begin{figure}
\center
\includegraphics[scale=0.5]{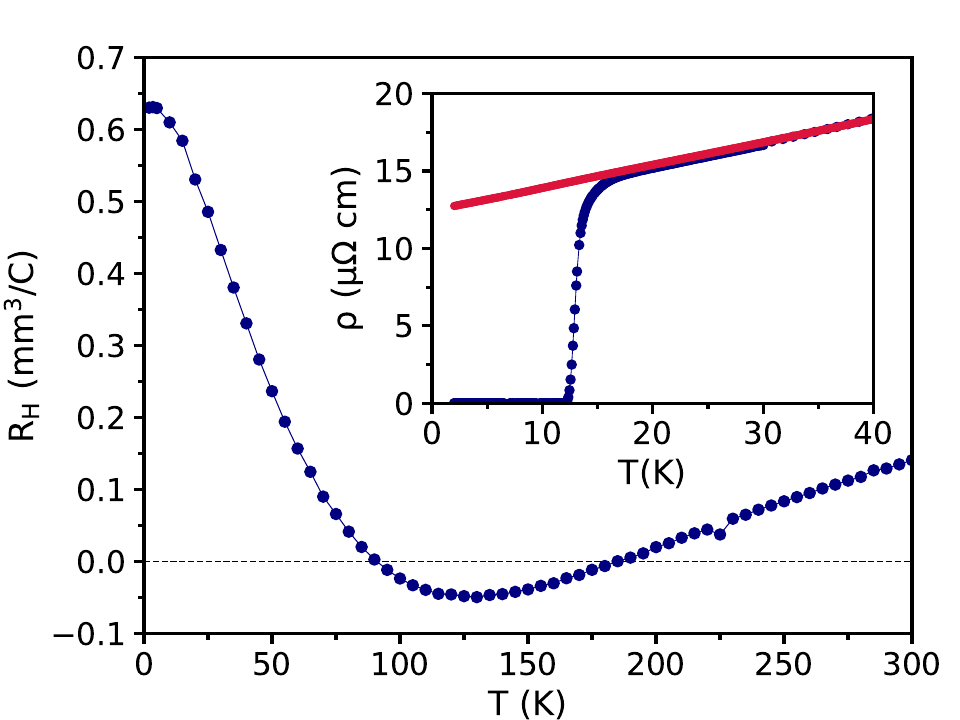}\\
\caption{The Hall coefficient as a function of temperature for \ce{PCCO} with $x=0.18$. $R_\mathrm{H}$ crosses zero at two different temperatures, $T_\mathrm{cr1} \cong~$\SI{91}{\kelvin} and $T_\mathrm{cr2} \cong~$\SI{180}{\kelvin}. The solid line is a guide to the eye. Inset: Resistivity $\rho (T)$ as a function of temperature at zero field (solid blue circle) and \SI{9}{\tesla} (straight red line).}
\label{1}
\end{figure}

Sign changes can also be observed for a fixed doping as temperature is varied in electron-doped cuprates. The best example of such behavior is that of overdoped \ce{Pr_{2-x}Ce_{x}CuO_{4}} (\ce{PCCO}) with $x = 0.17$ or $0.18$ for which two sign changes in $R_\mathrm{H}(T)$ are observed at two different temperatures \cite{dagan2016fermi,charpentier2010antiferromagnetic,fournier1998insulator,jin2011link} as is shown in Fig~\ref{1}. Although the low temperature Fermi surface for $x > 0.17$ is a unique large hole cylinder from ARPES \cite{matsui2007evolution} as antinodal and nodal arcs are now connected, the transport properties clearly indicate the persisting presence of electron-like carriers at intermediate temperatures. This inconsistency between transport and ARPES prompts further investigation into the underlying mechanisms governing the temperature dependence of transport in electron-doped cuprates, and more generally in all cuprates. Strong correlations can still lead to electron-like behavior for some of the states on this hole like FS according to Kontani \textit{et al.} \cite{kontani1999hall}. Using current vertex corrections (CVC) for a nearly antiferromagnetic (AF) Fermi liquid, Kontani \textit{et al.} showed that strong AF fluctuations can lead to an extra contribution to the current density no longer perpendicular to the FS, with the maximum impacts close to the crossing points of the FS with the antiferromagnetic Brillouin zone (AFBZ) limits, i.e., for wavevectors in proximity to the hot spots \cite{kontani1999hall}. These temperature-dependent corrections that are becoming important at intermediate temperatures give rise to a negative contribution to the Hall resistivity large enough to exceed the positive contribution from other parts of the FS leading to sign changes of $R_\mathrm{H}(T)$ as observed in Fig.~\ref{1}. In the studies by Jenkins \textit{et al.}  \cite{jenkins2009terahertz,jenkins2010origin}, Fermi-liquid CVC effects were identified as the potential source of the observed anomalous ac Hall transport in the THz regime for electron-doped cuprates, while for underdoping, the presence of small electron Fermi pockets and persistent negative Hall mass were attributed to CVC effects arising from antiferromagnetic fluctuations \cite{kontani1999hall}.

Overall, interactions split up the hole like FS into portions that continue to behave like holes and transform others that behave like electrons in applied electric and magnetic fields, leading to a situation very similar to a noninteracting electronic system with two types of carriers. Here, instead of having well-defined pockets, the carrier densities will be determined by the proportion of the whole FS with a characteristic behavior, hole- or electron-like, while the mobilities will represent averages over $\vec{k}$ space of their contributions to current. In what follows, we will assume for simplicity that the two-carrier model applies for all doping (below and above $x^*$). Eventually, one will need to connect the parameters extracted from the two-carrier model to the parameters of a theory taking into account interactions such as that presented in Ref. \cite{kontani1999hall}. Thus, we are expecting this crude two-carrier model based on parameters averaged over $\vec{k}$ space to fail in describing perfectly the measured electrical transport coefficients but capturing the essential features.

%%%%%%%%%%%%%%%%%%%%%%%%%%%%%%%%%%%%%%%
% Figure 1  : RH vs T                 %
%%%%%%%%%%%%%%%%%%%%%%%%%%%%%%%%%%%%%%%

%%%%%%%%%%%%%%%%%%%%%%%%%%%%%%%%%%%%%%%

Using the two-carrier model, one can examine specific predictions that are not commonly explored in usual transport measurements. For example, at $T_\mathrm{cr1} \cong~$\SI{91}{\kelvin}, the temperature at which $T_\mathrm{H}(T_\mathrm{cr1}) = 0$ for \ce{PCCO} with $x=0.18$ in Fig.~\ref{1}, higher-order field dependence of the Hall resistivity should be discernible according to the two-carrier model. In fact, around $T_{\mathrm{cr1}}$, the Hall resistivity can be written as a power series of the applied field $B$ in the limit of small fields, $\rho_{\mathrm{yx}}=R_{H} B+C B^{3}+ D B^{5}+. . .$ where $R_{\mathrm{H}} \sim 0$ at $T_{\mathrm{cr1}}$, but where $C$ and the following terms in the series are nonzero. Within a two-carrier model, these higher-order terms can be deduced as a function of the usual four parameters of the model, namely the density of holes and electrons, $p$ and $n$, and their respective mobilities $\mu_{\mathrm{p}}$ and $\mu_{\mathrm{n}}$. 

In this paper, we explore the presence of the nonlinear components of the Hall resistivity in \ce{PCCO} with $x=0.18$ and examine if the $B^{3}$ term is consistent with the prediction of the two-carrier model and data obtained from ARPES. As expected, this nonlinear term is easily observable at the temperature where the Hall coefficient is zero around $T_{\mathrm{cr1}} \sim~$\SI{91}{\kelvin} as observed in Fig.~\Ref{1}. We show that this $B^{3}$ term persists and grows at low temperature. Using the temperature dependence of the measured resistivity and the Hall coefficient, and assuming fixed values for the electron and hole densities deduced from ARPES, we can extract the temperature dependence of the two other parameters of the two-carrier model, namely the electron and hole mobilities. We then demonstrate that this particular set of densities and mobilities allows one to confirm the magnitude of the $B^{3}$ term observed experimentally, showing a semiquantitative agreement with a simplistic model for such a complex strongly correlated metal.

%%%%%%%%%%%%%%%%%%%%%%%%%%%%%%%%%%%%%%%
\section{Two-carrier model for nonlinear Hall resistivity}

Within the two-carrier model \cite{Sondheimer1948}, only four parameters are required to fully simulate the temperature dependence of the resistivity and the Hall coefficient: the electron and the hole densities, $n$ and $p$, and their corresponding mobilities $\mu _{\mathrm{n}}$ and $\mu _{\mathrm{p}}$. One needs to determine them for all temperatures. From the semiclassical Drude model, the conductivity of each type of carriers is given by $\sigma_\mathrm{n} = n e \mu_{\mathrm{n}}$ for electrons and $\sigma_{\mathrm{p}} = p e \mu_{\mathrm{p}}$ for holes \cite{ashcroft2010solid,nakamura1998physical,livanov1999hall}. In turn, the mobilities are then given by $\mu_{i} = \vert \frac{q_{i} \tau_{i}}{m_{i}^{*}} \vert$ where $i = n$ or $p$ while $\tau_{i}$ and $m_{i}$ are the relaxation time and the effective mass. Considering that the in-plane conductivity is isotropic ($\sigma _{\mathrm{xx}} = \sigma_{\mathrm{yy}}$ for electron-doped cuprates), the in-plane conductivity matrix when a magnetic field is applied along the c axis is given by:

\begin{align}
\tilde{\sigma} = \begin{pmatrix}
\sigma_{\mathrm{xx}} & \sigma_{\mathrm{B}}\\
-\sigma_{\mathrm{B}} & \sigma_{\mathrm{xx}}
\end{pmatrix}
\label{sigma}
\end{align}
where the diagonal and off-diagonal elements correspond to the longitudinal and transverse conductivity, respectively. This matrix is the sum $\tilde{\sigma} = \tilde{\sigma}_{\mathrm{n}} + \tilde{\sigma}_{\mathrm{p}}$ of the contributions from each carrier which are then given by:

\begin{align}
\tilde{\sigma}_{\mathrm{n}} = \dfrac{en \mu_{\mathrm{n}}}{1+\mu^{2}_{\mathrm{n}}B^{2}} \begin{pmatrix}
1 & -\mu_{\mathrm{n}}B\\
\mu_{\mathrm{n}}B & 1
\end{pmatrix}
\label{sigman}
\end{align}
for electrons and:

\begin{align}
\tilde{\sigma}_{\mathrm{p}} = \dfrac{ep \mu_{\mathrm{p}}}{1+\mu^{2}_{\mathrm{p}}B^{2}} \begin{pmatrix}
1 & \mu_{\mathrm{p}}B\\
-\mu_{\mathrm{p}}B & 1
\end{pmatrix}
\label{sigmap}
\end{align}
for holes. 

Since resistivity is the inverse of conductivity $\tilde{\rho} =\tilde{\sigma}^{-1} $, one can deduce the components of the resistivity matrix. This leads to the following expressions for the longitudinal resistivity and the Hall resistivity:

\begin{align}
\rho_{\mathrm{xx}}=\frac{1}{e}\frac{\left( p\mu_{\mathrm{p}}+n\mu_{\mathrm{n}} \right)+\mu_{\mathrm{p}}\mu_{\mathrm{n}} \left( n\mu_{\mathrm{p}}+p\mu_{\mathrm{n}} \right)B^{2}}{\left( p\mu_{\mathrm{p}}+n\mu_{\mathrm{n}} \right)^{2} +\mu_{\mathrm{p}}^{2} \mu_{\mathrm{n}}^{2} \left( p+n \right)^{2} B^{2}}
\label{rhoxxfull}
\end{align}
and,

\begin{align}
\rho_{\mathrm{yx}}=\frac{B}{e}\frac{\left( p\mu_{\mathrm{p}}^{2}-n\mu_{\mathrm{n}}^{2} \right)+\mu_{\mathrm{p}}^{2}\mu_{\mathrm{n}}^{2} \left( p-n \right) B^{2}}{\left( p\mu_{\mathrm{p}}+n\mu_{\mathrm{n}} \right)^{2}+\mu_{\mathrm{p}}^{2}\mu_{\mathrm{n}}^{2} \left( p+n \right)^{2} B^{2}}
\label{rhoyxfull}
\end{align}

These equations can be used to fit directly the data leading to the determination of the four free parameters $n$, $p$, $\mu_{\mathrm{n}}$ and $\mu_\mathrm{p}$ (see for example Refs. \cite{Y124twocarriers,Singha2017,Ishida2009}). In the low field limit when $\mu_\mathrm{n}B$ and $\mu_\mathrm{p}B \ll 1$, one can expand $\rho_\mathrm{xx}$ and $\rho_\mathrm{yx}$ in powers of the field as follows:

\begin{align}
\rho_{\mathrm{xx}} = \dfrac{1}{e} \dfrac{1}{n\mu_{\mathrm{n}}+p\mu_{\mathrm{p}}}+\dfrac{1}{e}\dfrac{np \mu_{\mathrm{n}}\mu_{\mathrm{p}}(\mu_{\mathrm{n}}+\mu_{\mathrm{p}})^{2}}{(n\mu_{\mathrm{n}}+p\mu_{\mathrm{p}})^{3}}B^{2}+...
\label{rhoxx}
\end{align}
and,

\begin{equation}
\begin{aligned}
\rho_{yx} = \dfrac{p\mu_{p}^{2}-n\mu_{n}^{2}}{e(n\mu_{n}+p\mu_{p})^{2}}B + \dfrac{np(p-n)\mu_{n}^{2}\mu_{p}^{2}(\mu_{n}+\mu_{p})^{2}}{e(n\mu_{n}+p\mu_{p})^{4}}B^{3} + \\ \dfrac{np(n-p)^{3}\mu_{n}^{4}\mu_{p}^{4}(\mu_{n}+\mu_{p})^{2}}{e(n\mu_{n}+p\mu_{p})^{6}}  B^{5} +...
\label{RH_Eq}
\end{aligned}
\end{equation}

In Eq.~\ref{RH_Eq}, the prefactors to the powers of $B$ in the Hall resistivity ($\rho_\mathrm{yx}=R_{\mathrm{H}} B+C B^{3}+D B^{5}+...$) can thus be all expressed in terms of $n$, $p$, $\mu_\mathrm{n}$ and $\mu_\mathrm{p}$ based on the two-carrier model.  
The usual linear Hall effect is defined as the first term ($R_\mathrm{H}$) while nonlinear contributions of higher orders in magnetic field are expected and should be most visible when the first term is zero, i.e, when $R_\mathrm{H}=0$. In this study, we extract experimentally the parameter $C$ of the $B^{3}$ contribution to the Hall resistivity at $T_\mathrm{cr1}$ and analyze it first qualitatively based on its expected value given by:

\begin{align}
C = \dfrac{np(p-n)\mu_{\mathrm{n}}^{2}\mu_{\mathrm{p}}^{2}(\mu_{\mathrm{n}} + \mu_{\mathrm{p}})^{2}}{e(n \mu_{\mathrm{n}} + p \mu_{\mathrm{p}})^{4}}  
\label{eq:C}
\end{align}

We follow with an approach to extract $C$ at other temperatures for which $R_{\mathrm{H}} \neq 0$. Finally, using the resistivity and the Hall effect measurements, we determine $\mu_{\mathrm{n}} (T)$ and $\mu_{\mathrm{p}} (T)$ and show that the four parameters $p$, $\mu_{\mathrm{p}}$, $n$ and $\mu_{\mathrm{n}}$ of the two-carrier model can be used to compute the temperature dependence of $C(T)$. The semiquantitative agreement is confirming the applicability of the two-carrier model for this overdoped electron-doped cuprate despite being a simplified description of the characteristics of the electronic structure driven by strong electronic interactions and the related carrier dynamics. 

%%%%%%%%%%%%%%%%%%%%%%%%%%%%%%%%%%%%%%%
\section{Experiments and methods}

Thin films of overdoped $c$ axis oriented \ce{PCCO} ($x = 0.18$) are deposited from an off-stoichiometric target with excess \ce{CuO} \cite{roberge2009improving} onto ($001$)-oriented \ce{LSAT} substrates using a \SI{248}{\nm} KrF excimer laser with a repetition rate of \SI{10}{\Hz}.  The substrate temperature is maintained at \SI{850}{\degreeCelsius} in a gas mixture of \SI{200}{mTorr} of \ce{N_{2}O} and \SI{150}{mTorr} of \ce{O_{2}} environment during the entire deposition process. The typical thickness of the films is around \SI{130}{\nm}. Right after deposition, \textit{in-situ} annealing in vacuum is performed at the same temperature for \SI{10}{\min} before quenching to room temperature. The crystalline structure of the layers are characterized using a Bruker D8 Discover high-resolution x-ray diffractometer in the $2 \theta - \omega$ configuration. The results confirm the high quality of the samples similar to previous reports \cite{roberge2009improving,charpentier2010antiferromagnetic}. \\
To perform accurate electrical transport measurements, Hall bars were fabricated using three different approaches. As the electron-doped cuprates are sensitive to chemical and physical etching, Hall bars were defined by photolithography and films were etched by ion milling on a liquid-nitrogen cooled stage. As a second approach, Hall channels were drawn using a fine diamond scriber tip. Finally, we developed substrate selective technique similar to that of Ref. \cite{Morales2005} that allows us to avoid the eventual degradation of the materials. In a nutshell, an amorphous layer of \ce{SiO_{2}} or \ce{Al_{2}O_{3}} is deposited on a masked substrate. After lift-off, the oxide thin film is deposited on the exposed substrate area defining a Hall bar. The resulting epitaxial thin films on the unmasked area is ready to be measured with no extra processing except for the application of contacts avoiding the potential degradation of post-deposition processes \cite{Ghotb2024}. The three separate approaches lead to the same resistivity and Hall effect. The longitudinal and Hall resistivity measurements are carried out using a Physical Property Measurement System (PPMS) from Quantum Design wih a magnetic field applied perpendicular to the sample surface up to \SI{9}{\tesla} in the temperature range from \SIrange[range-units=single]{2}{300}{\kelvin}. For the sensitive Hall voltage measurements, an external current source and nanovoltmeters are used to improve the signal to noise ratio. In this specific case, the Hall (transverse) voltage is measured at a fixed temperature as a function of positive and negative applied magnetic field. Special care is taken to stabilize the field prior to measurements and the Hall resistivity is computed to eliminate any remaining offset due to the misalignment of the Hall contacts. For the proper extraction of the nonlinear contribution to $\rho_{\mathrm{yx}}$, the antisymmetric Hall voltage is determined by the subtraction of $\rho_{\mathrm{yx}}$ for positive and negative fields [$\rho_{\mathrm{yx}}(B)=(\rho_{\mathrm{yx}}(+B)-\rho_{\mathrm{yx}}(-B))/2$].

%%%%%%%%%%%%%%%%%%%%%%%%%%%%%%%%%%%%%%%
\section{Results and discussion}

The inset in Fig~\ref{1} shows the $ab$-plane resistivity as a function of temperature for a \ce{PCCO} film under zero and \SI{9}{\tesla} applied magnetic fields. The transition temperature is found to be \SI{12.4}{\kelvin} from the peak in the derivative of the resistivity with respect to the temperature at zero magnetic field. The residual resistivity ($\rho_{0}$) is estimated to be \SI{14}{\mu \ohm~\cm} from the extrapolation of the normal state data at \SI{9}{\tesla} which compares well with other reports \cite{higgins2006role,dagan2004evidence,charpentier2010antiferromagnetic,tafti2014nernst}. The normal state in \ce{PCCO} ($x=0.18$) exhibits strange metal transport which is manifested as a linear-in-T resistivity from $T=0$ to \SI{60}{\kelvin} and $\sim T^{2}$ resistivity from \SIrange[range-units=single]{40}{300}{\kelvin} resembling the previous reports for this material at $x=0.17$ \cite{fournier1998insulator,jin2011link,legros2019universal}.
The absence of any upturn in the normal state resistivity confirms that our sample is above the critical doping ($x \cong 0.165$) \cite{dagan2004evidence,tafti2014nernst,armitage2010progress,charpentier2010antiferromagnetic}. 

The Hall coefficient as a function of temperature is also shown in Fig.~\ref{1}. It is determined from the measurements of the transverse resistivity in the temperature range from \SIrange[range-units=single, range-phrase=--]{2}{300}{\kelvin} under a large magnetic field of \SI{\pm 9}{\tesla}. $R_\mathrm{H}$ exhibits a strong temperature dependence as is commonly observed in cuprates. As underlined previously \cite{armitage2010progress,charpentier2010antiferromagnetic,dagan2004evidence,fournier1998insulator,greene2020strange}, the most intriguing feature of the Hall coefficient in overdoped electron-doped cuprates is the sign reversals in $R_{\mathrm{H}}$ in the normal state, as it crosses zero at $T_{\mathrm{cr1}} \cong~$\SI{92}{\kelvin} and $T_{\mathrm{cr2}} \cong~$\SI{180}{\kelvin} for our \ce{PCCO} $x=0.18$ thin films. This behavior has also been observed with various values of $T_{\mathrm{cr}}$ for other cerium contents in \ce{PCCO} ($x = 0.16$, $0.17$), close to optimal doping ($x=0.15$) in \ce{NCCO} and between $0.13$ and $0.14$ for \ce{LCCO} \cite{dagan2016fermi,charpentier2010antiferromagnetic,jiang1994anomalous,fournier1997thermomagnetic,sawa2002electron,sarkar2017fermi}  and represents a common trend in this family. The great sensitivity of the values of $T_{\mathrm{cr}}$ on sample preparation (reduction/oxygen stoichiometry) and cerium content is noticeable. As shown below, exploring nonlinear Hall resistivity may be a great tool to extract more information on the sources of variability on the electronic properties of these various samples.

\begin{figure}
\center
\includegraphics[scale=0.45]{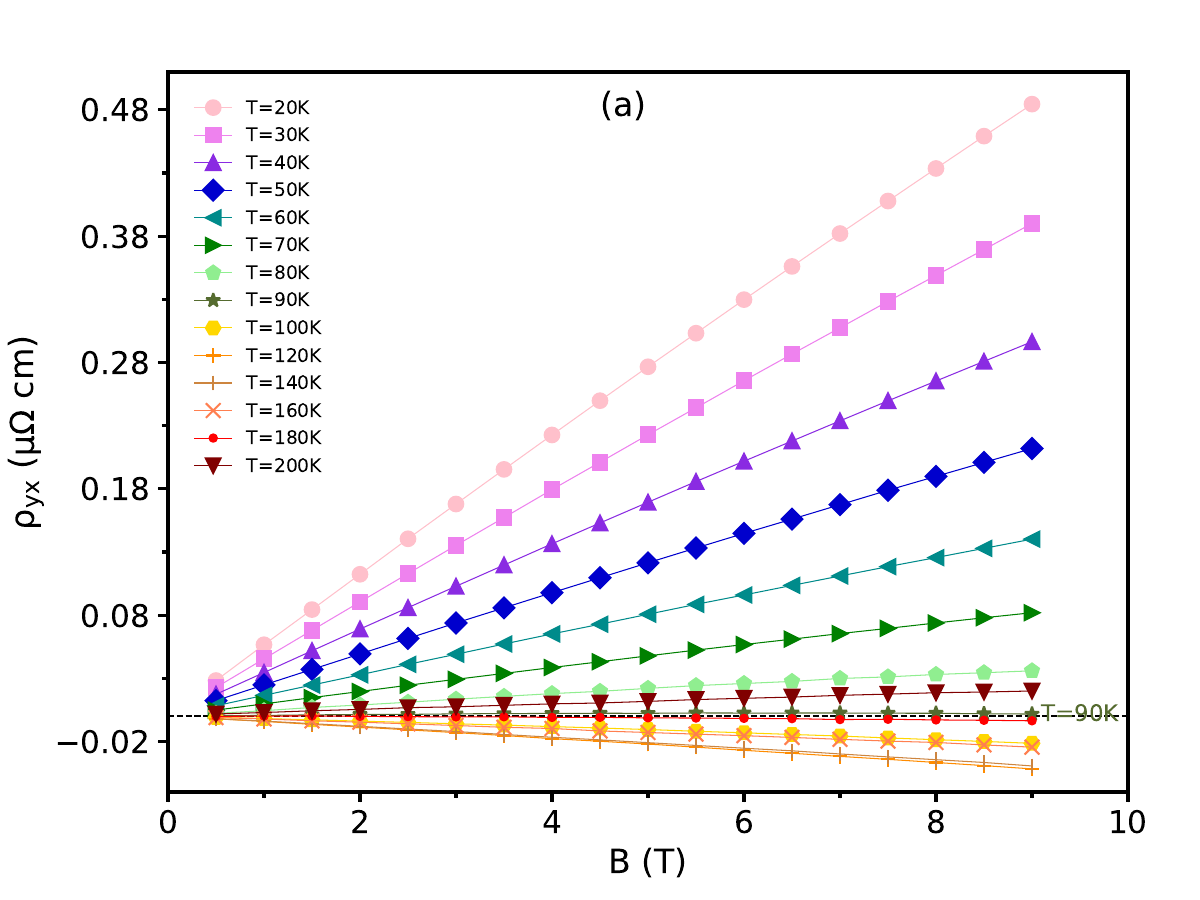}
\includegraphics[scale=0.45]{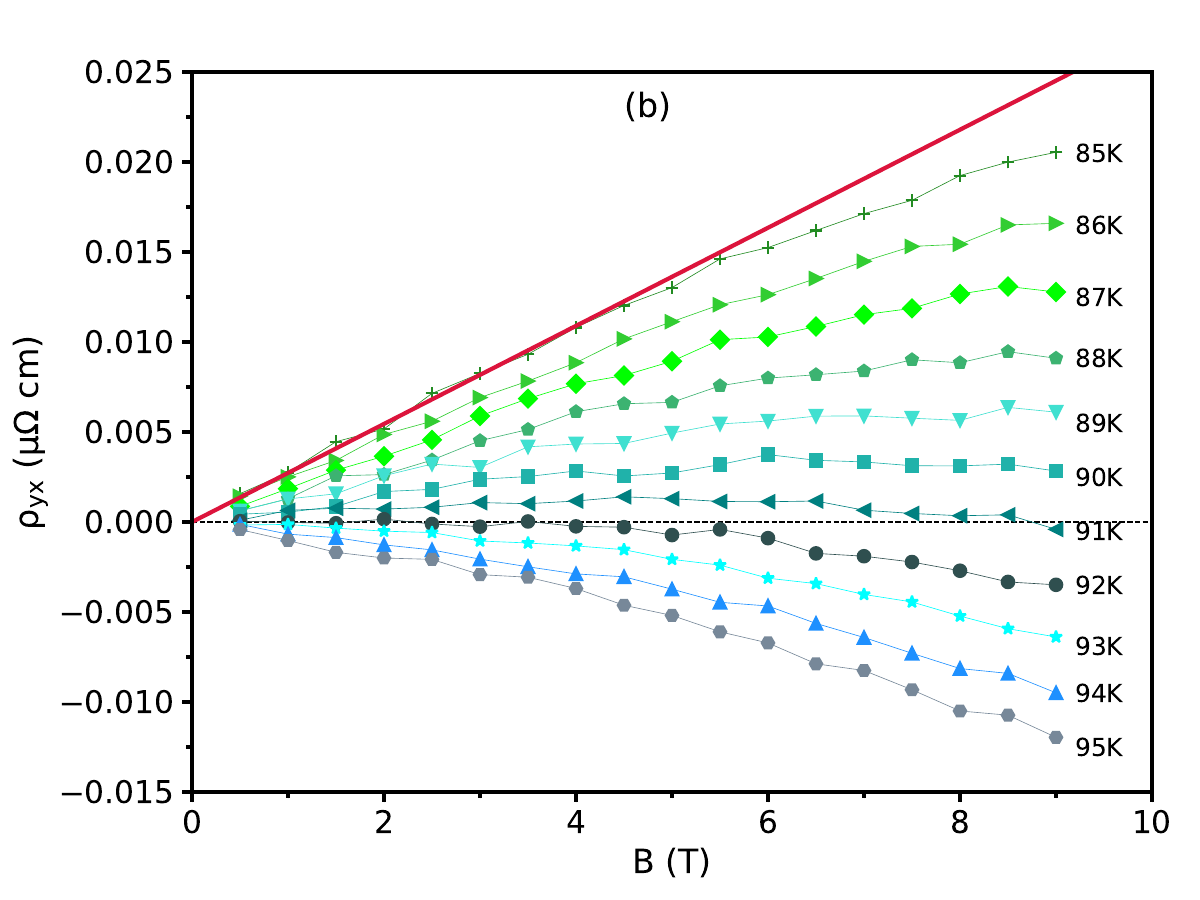}
\caption{(a) Isothermal Hall resistivity ($\rho_{\mathrm{yx}}$) as a function of magnetic field in the temperature range of \SIrange[range-units=single, range-phrase=--]{20}{200}{\kelvin}. Solid lines are guides to the eye. (b) Isothermal Hall resistivity as a function of magnetic field at temperature increments of \SI{1}{\kelvin} in the vicinity of the low-T crossing point at $T_{\mathrm{cr1}} \cong ~$\SI{91}{\kelvin}. The solid red line serves as a straight line reference, highlighting deviations from linearity. Solid lines are guides to the eye. }
\label{2}
\end{figure}

According to the first term of the two-carrier model in Eq.~\ref{RH_Eq}, the Hall coefficient is zero at $T_{\mathrm{cr1}}$ when $p \mu_{\mathrm{p}}^{2} = n \mu_{\mathrm{n}}^{2}$. Equation.~\ref{RH_Eq} predicts also that higher-order field dependencies in $\rho_{\mathrm{yx}}(B)$ should be nonzero. Figure~\ref{2}(a) displays the Hall resistivity isotherms measured in the temperature range of \SIrange[range-units=single, range-phrase=--]{20}{200}{\kelvin} up to the magnetic field of \SI{9}{\tesla}. In agreement with the $R_\mathrm{H}(T)$ data in Fig.~\ref{1}, the $B \rightarrow 0$ slope of $\rho_{\mathrm{yx}}(B)$ is changing from positive to negative at $T_{\mathrm{cr1}} \cong ~$\SI{91}{\kelvin}, reaches a maximum negative value around \SI{120}{\kelvin} and then becomes positive again beyond $T_{\mathrm{cr2}} \cong ~$\SI{180}{\kelvin}. 

%%%%%%%%%%%%%%%%%%%%%%%%%%%%%%%%%%%%%%%
% Figure 2  : Nonlinear rhoyx        %
%%%%%%%%%%%%%%%%%%%%%%%%%%%%%%%%%%%%%%%
%%%%%%%%%%%%%%%%%%%%%%%%%%%%%%%%%%%%%%%

To further reveal the nonlinearity of $\rho_{\mathrm{yx}}(B)$, a subset of the isotherms measured with a temperature increment of \SI{1}{\kelvin} in the vicinity of the low-T crossing point at $T_{\mathrm{cr1}} \cong ~$\SI{91}{\kelvin} is shown in Fig.~\ref{2}(b). The nonlinearity becomes most apparent as expected near this crossing point as the slope at $B\rightarrow 0$ of these isotherms is almost zero. At \SI{91}{\kelvin} for instance, $\rho_{\mathrm{yx}}(B)$ is zero at a field (\SI{9}{\tesla}) used to evaluate the Hall coefficient in Fig.~\ref{1}, but $R_H$ would not be exactly zero if computed at an intermediate field (say \SI{5}{\tesla}). At \SI{91}{\kelvin}, $\rho_{\mathrm{yx}}(B)$ still shows a nonzero $B\rightarrow 0$ slope defining the real Hall coefficient as the first term in Eq.~\ref{RH_Eq}. In fact, the linear term (the slope) is really zero at $\sim ~$\SI{92}{\kelvin} according to Fig.~\ref{2}(b). Although this linear term quickly dominates the field dependence away from \SI{91}{\kelvin}, one notices the persistence of the extra field dependence at \SI{85}{\kelvin} and \SI{95}{\kelvin} in Fig.~\ref{2}(b) (the $\rho_{\mathrm{yx}}(B)$ traces are not straight lines). Moreover, this extra field dependence results in a negative curvature for all temperatures implying that $C$ in Eq.~\ref{eq:C} is negative. In fact, a close inspection of $\rho_{\mathrm{yx}}(B)$ in Fig.~\ref{2}(a) reveals the persistence of this negative extra term for all temperatures down to \SI{20}{\kelvin}.

Nonlinear Hall resistivity in \ce{PCCO} has already been reported by Li \textit{et al.} \cite{li2007high} in the very high-field regime compared to ours, as large as \SI{58}{\tesla} using pulsed magnetic field. Their observations of a large positive curvature in $\rho_{\mathrm{yx}}(B)$, combined to the change from a quadratic to a linear field regime in the magnetoresistance, led the authors to propose a spin-density wave-induced FS reconstruction as the possible origin of these high-field manifestations. We are not debating these results and the related interpretation as we are focusing our attention to much lower magnetic field range than in Ref. \cite{li2007high} with a maximum of \SI{9}{\tesla}. The effect we observe is also much weaker than that observed in Ref. \cite{li2007high} and may not be related directly (not the same origin).   

From Eq.~\ref{eq:C}, a negative value for $C$ implies that $n > p$. This is a straightforward result of our two-carrier analysis. It signifies that, despite observing a hole like cylinder from ARPES for $x > x^{*} \cong 0.165$ \cite{matsui2007evolution}, most of the current in this material is dominated by carriers behaving as electrons at temperatures around $T_{\mathrm{cr1}}$. Moreover, since the Hall coefficient is zero at \SI{92}{\kelvin} and thus $p \mu_{\mathrm{p}}^{2} = n \mu_{\mathrm{n}}^{2}$, $n > p$ implies that $\mu_{\mathrm{p}} > \mu_{\mathrm{n}}$. Although the hole density is low compared to electron-like carriers according to the two-carrier model, their high mobility allows them to dominate at some specific temperatures, namely at very low and at high temperatures, while the electrons dominate at intermediate temperatures. 

These observations can be reconciled to some extent with recent ARPES data that estimate the proportion of electron to hole densities for the same family for $x$ values lower, but very close to $x^{*}$ in \ce{NCCO} \cite{he2019fermi}. He \textit{et al.} show that the ratio of the densities of electrons to holes should be roughly $n/p \cong 5 - 6$. As the FS arcs merge at $x^{*} \cong 0.165$ as a result of the collapsing pseudogap, the proportion that these arcs occupy on the large hole cylinder should remain pretty similar explaining in parts our experimental conclusion that $n > p$ for $x = 0.18$, just above $x^{*}$. Of course, this interpretation is valid if a mechanism like that proposed by Kontani \textit{et al.}  \cite{kontani1999hall} modifies the dynamics of electrons on portions of the FS explaining the negative Hall effect for $x$ beyond $x^{*}$ despite the large hole like cylinder \cite{kontani1999hall}.

\begin{figure}
\center
\includegraphics[scale=0.48]{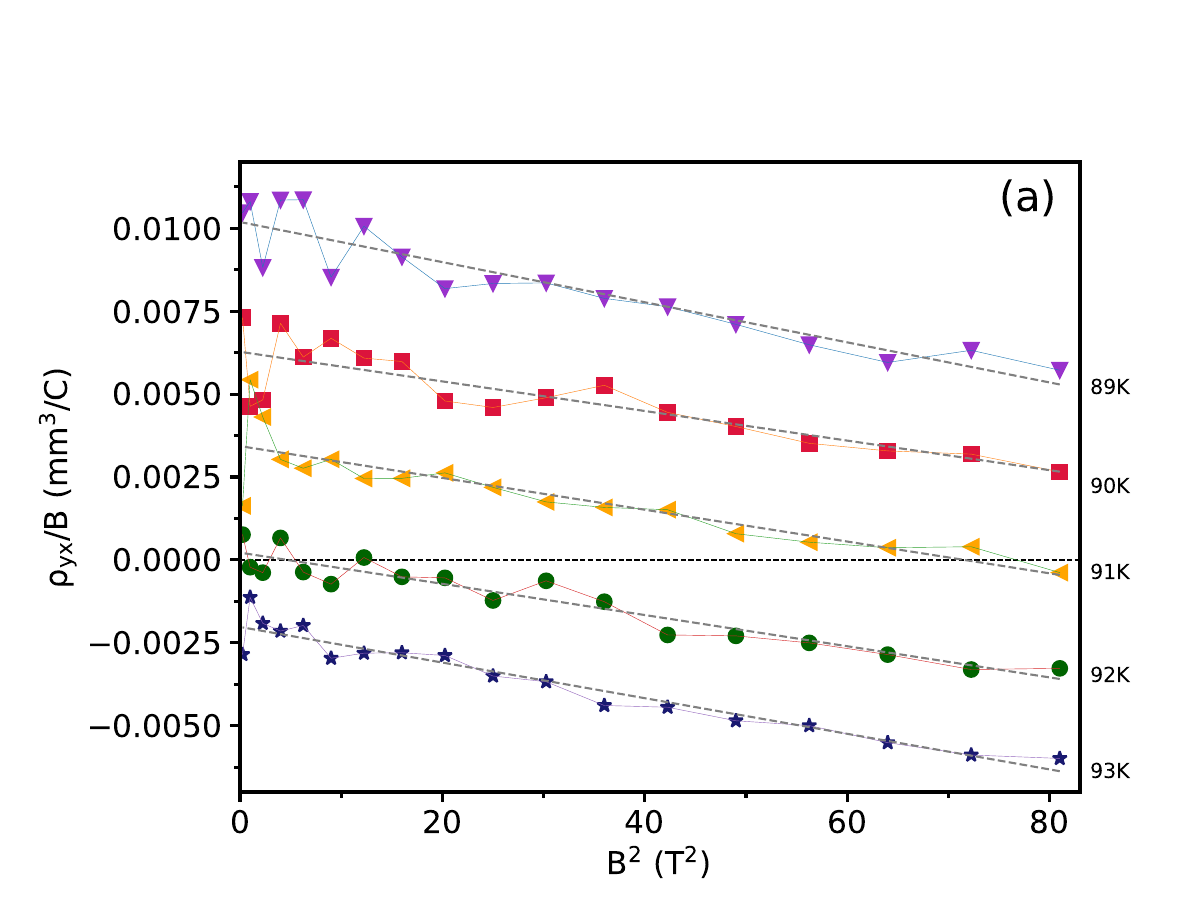}
\includegraphics[scale=0.48]{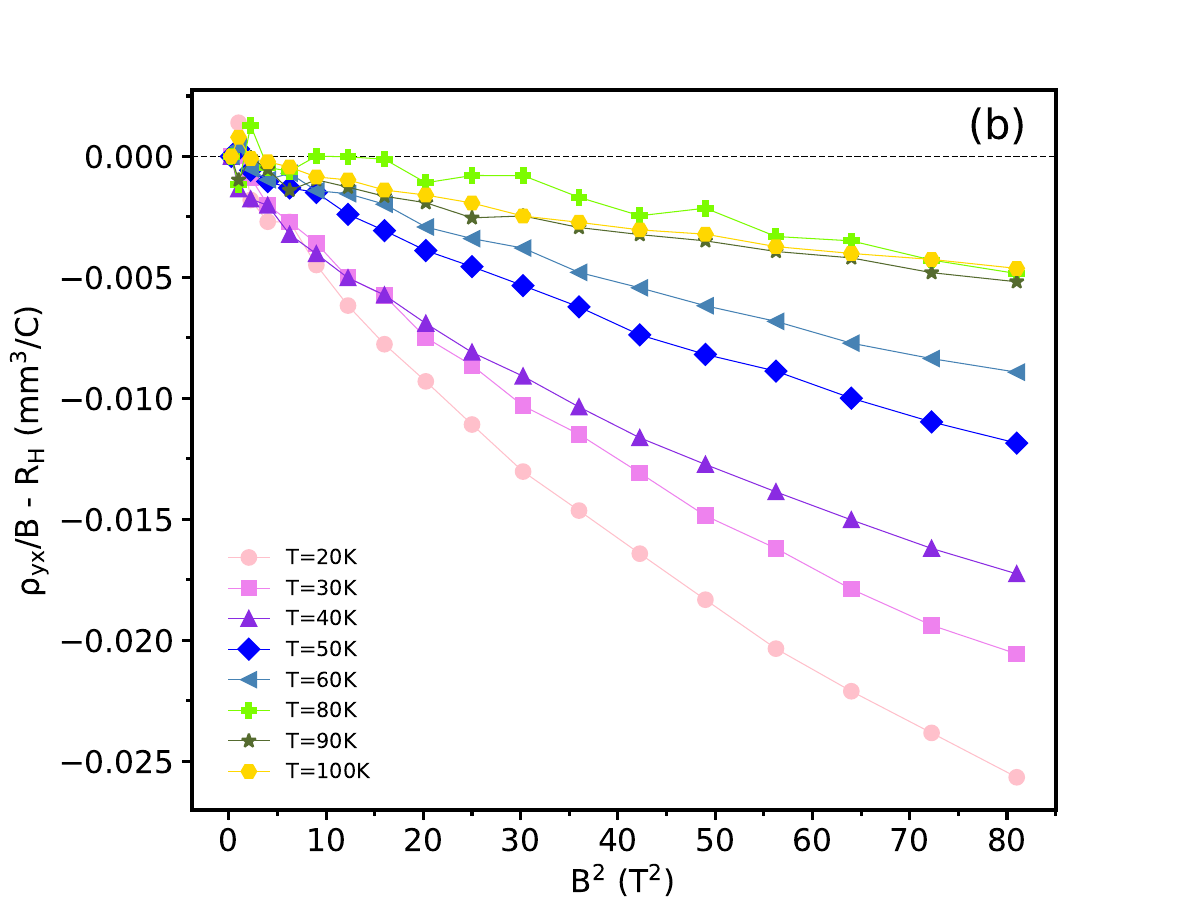}
\caption{(a) $\rho_{\mathrm{yx}}/B$ as a function of $B^{2}$, showing a negative slope in the temperature range of \SIrange[range-units=single, range-phrase=--]{85}{95}{\kelvin}. The extrapolated value at $B=0$ is $R_{\mathrm{H}}(0)$, the first term in Eq.~\ref{RH_Eq} while the slope is the coefficient of the second term in the same equation.  (b) $\rho_{\mathrm{yx}}/B-R_{\mathrm{H}}$ as a function of $B^{2}$ from \SIrange[range-units=single]{20}{140}{\kelvin}.}
\label{3}
\end{figure}

Based on Eq.~\ref{RH_Eq}, the coefficient of the $B^{3}$ term of $\rho_{\mathrm{yx}}(B)$ can be extracted by plotting $\rho_{\mathrm{yx}}/B$ as a function of $B^{2}$ as shown in Fig~\ref{3}(a) first for isotherms measured near the crossing point at \SI{91}{\kelvin} (see Fig.~\ref{2}(b)). The slope of $\rho_{\mathrm{yx}}/B$ as a function of $B^{2}$ is negative and its value can be estimated from a fit for each isotherm. Figure.~\ref{3}(b) presents $\rho_{\mathrm{yx}}/B$ as a function of $B^{2}$ away from the crossing point. In this specific figure, we subtracted the first term ($R_{\mathrm{H}}$) from the data to present the curves at various temperatures on the same plot, helping us to appreciate the temperature dependence of the nonlinear contributions. In both figures, $\rho_{\mathrm{yx}}(B)/B$ would be simply a constant independent of field if the only term contributing in Eq.~\ref{RH_Eq} was $R_{\mathrm{H}}$. Despite the dominance of the linear term away from the crossing point, Fig.~\ref{3}(b) demonstrates that one can still track this nonlinear contribution to $\rho_{\mathrm{yx}}/B$ from \num{20} up to $\cong$\SI{140}{\kelvin} before the signal to noise ratio becomes too small. At the lowest temperatures (\num{10} to \SI{40}{\kelvin}), we observe additional \textit{positive} bending to the curves while it becomes purely $B^{3}$ beyond \SI{50}{\kelvin}. The additional low-temperature nonlinearity is probably a signature of the following term in the series in Eq.~\ref{RH_Eq}, namely the $B^5$ term that we are expecting to be positive when $n > p$. Another possible origin may be the spin-density wave-induced FS reconstruction proposed by Li \textit{et al.} \cite{li2007high}.  

%%%%%%%%%%%%%%%%%%%%%%%%%%%%%%%%%%%%%%%
% Figure 3  : rhoyx/B vs B^2          %
%%%%%%%%%%%%%%%%%%%%%%%%%%%%%%%%%%%%%%%

%%%%%%%%%%%%%%%%%%%%%%%%%%%%%%%%%%%%%%%

\begin{figure}
\center
\includegraphics[scale=0.5]{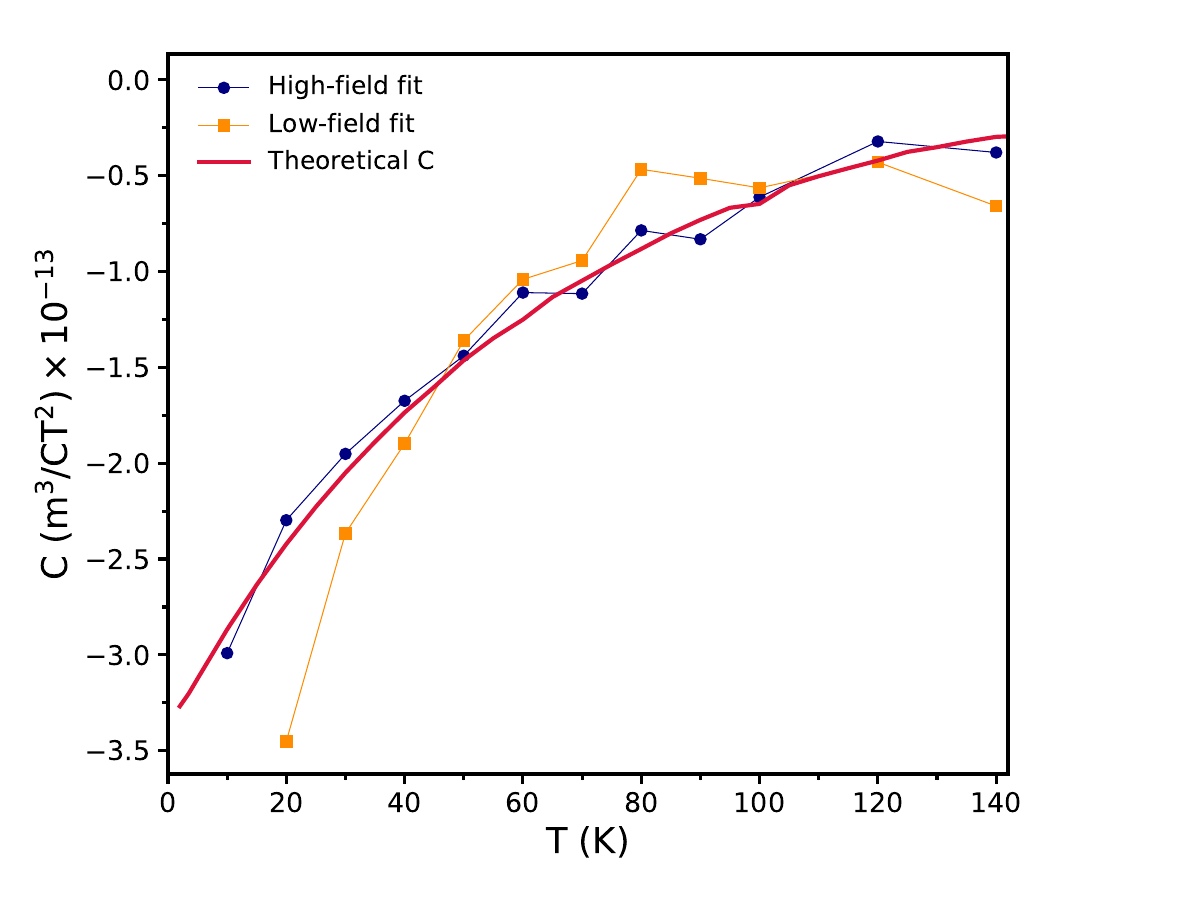}\\
\caption{Comparison of the experimental and the theoretical coefficient $C$ of the nonlinear field dependence of the Hall resistivity as a function of temperature. 
The solid circles represent experimental points extracted from the slopes of Fig.~\ref{3} with a high-field fit, joined by a guide to the eye.  The solid squares depict the C parameters obtained from a low-field fit (up to \SI{5}{\tesla}). The solid red line is computed using Eq.~\ref{eq:C} from the two-carrier model using $n =~$\SI[per-mode = symbol]{1.8e21}{\per \cm\cubed} and $p =~$\SI[per-mode = symbol]{0.3e21}{\per \cm\cubed} and the mobilities presented in Fig.~\ref{5} (see text). }
\label{4}
\end{figure}

Figure~\ref{4} presents the temperature dependence of the $B^{3}$ prefactor defined here as $C$. Due to the presence of the additional bending for the low temperature isotherms, the $C$ parameter is obtained through a low-field fit (roughly up to \SI{5}{\tesla}) of the curves in Fig.~\ref{3}.  The $C$ parameter remains negative over the whole temperature range and shows a strong temperature dependence up to \SI{140}{\kelvin}. This implies that $n > p$ for the same range of temperature, at least \cite{Comment}. According to Eq.~\ref{eq:C}, this parameter gets its temperature dependence entirely from the hole and electron mobilities if one assumes that $n$ and $p$ are independent of temperature.

%%%%%%%%%%%%%%%%%%%%%%%%%%%%%%%%%%%%%%%
% Figure 4  : C vs T                  %
%%%%%%%%%%%%%%%%%%%%%%%%%%%%%%%%%%%%%%%
%%%%%%%%%%%%%%%%%%%%%%%%%%%%%%%%%%%%%%%

\begin{figure}
\center
\includegraphics[scale=0.45]{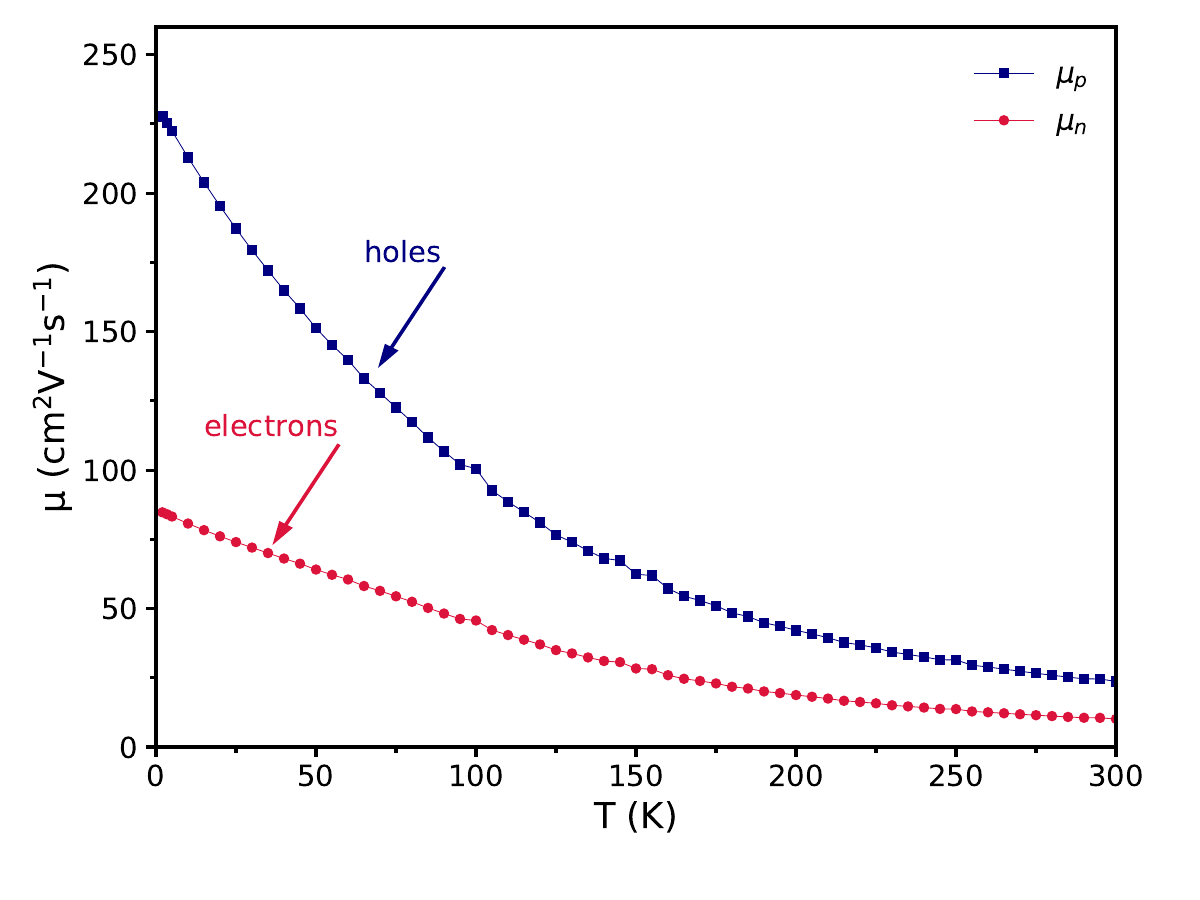}\\
\caption{Temperature dependence of the mobilities of holes (top) and electrons (bottom) extracted using the resistivity and the Hall coefficient based on the two-carrier model.}
\label{5}
\end{figure}

Using our measured $\rho(T)$ and $R_{\mathrm{H}}(T)$ and assuming, for simplicity, specific temperature-independent $n$ and $p$ values for the carrier densities of electrons and holes \cite{Comment}, one can extract the mobility of electrons ($\mu_{\mathrm{n}}$) and holes ($\mu_{\mathrm{p}}$) as a function of temperature based on the first terms of Eqs.~\ref{rhoxx} and ~\ref{RH_Eq}. For this purpose, we rely on the estimates of the electron and hole densities deduced from ARPES measurements on \ce{NCCO} samples \cite{matsui2005angle,he2019fermi} of $n =~$\SI[per-mode = symbol]{1.8e21}{\per \cm\cubed} and $p =~$\SI[per-mode = symbol]{0.3e21}{\per \cm\cubed}, leading to $n/p = 6$. Figure~\ref{5} presents the extracted electron and hole mobility as a function of temperature. The first noticeable result is the fact that the hole mobility is larger than the electron one by more than a factor of two for all temperatures. This is consistent with our conclusion above from our first analysis of the negative $C$ parameter ($n > p$) combined to the zero Hall coefficient at $T_{\mathrm{cr1}} =~$\SI{91}{\kelvin} implying that $\mu_{\mathrm{p}} > \mu _{\mathrm{n}}$. We also observe that the temperature dependence of both mobilities is rather weak, approaching $1/T$ only for temperatures above \SI{100}{\kelvin}. This is somehow a surprise as we were expecting power laws close to $1/T^{2}$, at least at high temperatures. Nevertheless, these values of  carrier mobility are consistent with the Hall mobility estimated for hole-doped cuprates using the Hall effect: $\mu_{\mathrm{H}} = R_{\mathrm{H}} / \rho_{\mathrm{xx}} \sim~$\SI{10}{\cm\squared\per\volt\per\second} optimally doped \ce{YBa_{2}Cu_{3}O_{7}} \cite{muHYBCO} and \ce{La_{1.85}Sr_{0.15}CuO_{4}} \cite{ono2007strong}. This also compares well with the Hall mobility of our \ce{PCCO} $0.18$ films which is found to be around \SI{50}{\cm\squared\per\volt\per\second} at \SI{2}{\kelvin}. The extracted values of $\mu_{\mathrm{n}}$ and $\mu_{\mathrm{p}}$ validate our assumption that $\mu_{\mathrm{n}}B$ and $\mu_{\mathrm{p}}B \ll 1$ used to justify the series expansions in Eqs.~\ref{rhoxx} and \ref{RH_Eq}. Finally, our results are in agreement with a previous study showing that electrons dominate the conductivity \cite{Dagan2007} since their density is assumed above to be $6$ times larger than the hole density while their mobility is at best $3$ times smaller than that of holes according to Fig.~\ref{5}.

%%%%%%%%%%%%%%%%%%%%%%%%%%%%%%%%%%%%%%%
% Figure 5  : mu_n and mu_p vs T      %
%%%%%%%%%%%%%%%%%%%%%%%%%%%%%%%%%%%%%%%
%%%%%%%%%%%%%%%%%%%%%%%%%%%%%%%%%%%%%%%

Using these values of $\mu_{\mathrm{p}}$ and $\mu_{\mathrm{n}}$ for all temperatures together with the selected values of $n$ and $p$, the expected coefficient $C$ of the $B^{3}$ term in the Hall resistivity can be computed also as a function of temperature from Eq.~\ref{eq:C}. Figure.~\ref{4} presents this computed $C$ parameter as the solid red line. We find a close agreement between the theoretical and the experimental values of $C$ at high temperatures while it deviates at low temperatures. Despite the obvious complexity of the electronic properties of this family of materials, the fact that the four parameters of a simplistic two-carrier model, $n$, $p$, $\mu_{\mathrm{n}}$ and $\mu_{\mathrm{p}}$, extracted from an analysis of $\rho (T)$ and $R_{\mathrm{H}}(T)$ can reproduce semiquantitatively the magnitude and the temperature dependence of the next leading term in the (nonlinear) field dependence of the Hall resistivity is likely a good indication that the two-carriers model can serve as a simple approach for analyzing the nonlinear Hall effect in electron-doped cuprates, for example as we may want to explore its doping dependence across the critical doping $x^{*} \cong 0.165$. 

For these overdoped electron-doped cuprates, a clear message from this interesting success of the two-carrier model is that large portions of the hole like Fermi surface of these materials lead to transport as if carriers were electrons. A more complete link between this two-carrier model and a theory that explains the origin of this electron-like behavior from a hole Fermi surface as that proposed by Kontani \textit{et al.} \cite{kontani1999hall} is necessary. Moreover, it would be interesting to see if other strongly correlated materials with similar sign reversal of their Hall coefficient show similar nonlinear Hall resistivity.  

%%%%%%%%%%%%%%%%%%%%%%%%%%%%%%%%%%%%%%%
\section{Conclusion}

In summary, we explored the nonlinear Hall resistivity of overdoped \ce{PCCO} cuprate at $x=0.18$ using the two-carrier model. The Hall coefficient of \ce{PCCO} undergoes two sign changes at temperatures \SIlist[list-units=single]{91;180}{\kelvin}. At \SI{91}{\kelvin}, we show that the Hall resistivity presents nonzero nonlinear contributions. This nonlinear effect is negative, it persists from $10$ to \SI{140}{\kelvin} and it is consistent with the expectations of the two-carrier model. In particular, we deduce that the density of electrons $n$ is higher than that of holes $p$ in agreement qualitatively with observations from ARPES for nearby doping and that the mobility of the holes $\mu_{\mathrm{p}}$ is larger than that of electrons $\mu_{\mathrm{n}}$. Using this two-carrier model, we show that the four parameters of the model, $n$, $p$, $\mu_{\mathrm{n}}$ and $\mu_{\mathrm{p}}$, can be deduced from measurements of the resistivity and the Hall coefficient as a function of temperature and they can be used in turn to compute the contribution of the nonlinear Hall resistivity, approaching closely the measured values at all temperatures. Although angle-resolved photoemission spectroscopy measurements suggest the presence of a large hole like cylindrical Fermi surface in overdoped electron-doped cuprates, the nonlinear Hall resistivity indicates that the dynamics of electrons on certain portions of the Fermi surface undergoes significant modifications due to strong interactions, causing these regions to exhibit electron-like behavior rather than holes.\\

%%%%%%%%%%%%%%%%%%%%%%%%%%%%%%%%%%%%%%%
\section*{Acknowledgment}

The authors thank S. Pelletier, B. Rivard,  M. Abbasi Eskandari,  P. Reulet, E. Blais and F. Naud for technical support. The authors also gratefully acknowledge Professor. André-Marie Tremblay for the helpful discussions. This work is supported by the Natural Sciences and Engineering Research Council of Canada (NSERC) under Grant No. RGPIN-2018-06656, the Canada First Research Excellence Fund (CFREF), the Fonds de Recherche du Québec - Nature et Technologies (FRQNT) and the Université de Sherbrooke. 

\bibliographystyle{apsrev4-2}
\bibliography{mybibfile.bib}

\end{document}